\documentstyle[preprint,aps]{revtex}
\begin{document}
\draft 
\preprint{IFT-P.022/97}
\title{ Short Distance QCD Contribution to the Electroweak 
Mass Difference of Pions}
\author{A.\ A.\ Natale*}
\footnotetext{*e-mail: natale@axp.ift.unesp.br}
\address{Instituto de F\'{\i}sica Te\'orica,
Universidade  Estadual Paulista, \\
Rua Pamplona 145, 01405-900 S\~ao Paulo, Brazil.}
\maketitle
\begin{abstract}
It is known that the short distance QCD contribution to the
mass difference of pions is quadratic on
the quark masses, and irrelevant with respect to the long
distance part. It is also considered in the
literature that its calculation contains infinities, which should
be absorbed by the quark mass renormalization.
Following a prescription by Craigie, Narison and Riazuddin,
of a renormalization group improved perturbation theory
to deal with the electromagnetic mass shift problem in QCD,
we show that the short distance QCD contribution to the 
electroweak pion mass difference (with $m_u = m_d \neq 0$) is 
finite and, of course, its value is negligible compared 
to other contributions. 
\end{abstract}


Recently there has been a lot of interest on the mass differences
of pions and kaons. There are many reasons for this. These mass
differences are important because they provide relations among 
the light quark masses, as well as they are a quite nice
example of strong isospin breaking. But most of the renewed interest 
lies mainly in the possibility of applying the latest 
techniques to handle the strong interaction physics of the
simplest hadrons. A very complete and detailed calculation of these mass
differences has been recently performed by Donoghue and 
P\'erez~\cite{don}, and this work contains most of the relevant 
references up to now on this subject.

As explained in Ref.~\cite{don}, the pion electromagnetic mass 
difference amplitude receives different contributions from the long
and short distance QCD, with the first one dominating the calculation,
and the second one canceling out at zeroth and first order in
the quark masses. The short distance QCD contribution to the
pion mass difference (proportional to the quark
masses) was not calculated in Ref.~\cite{don}. It
was correctly neglected because it is quadratic in the quark
masses. However, when we look to the many references discussing
this problem out of the chiral limit, {\it i.e.} taking 
into account a non-zero quark mass, which
is important for the kaon mass difference (see,
for instance, Ref.~\cite{bijnens}), we note the 
comment that this calculation gives an infinite result,
which must be renormalized by a quark mass counterterm. 
Actually, it is generally
agreed that the short distance QCD part of the pion mass
difference is small, and as far as we know we cannot find in the 
literature a precise numerical evaluation of its finite part.
It seems also that in the most recent 
papers about mass differences, the weak interaction contribution
has not been included and, although small, it
is as important as the electromagnetic
one when computing the high energy part of the mass splitting,
because it modifies the convergence of the calculation.

It is the purpose of this note to show, according to
a renormalization group improved perturbation theory prescription
by Craigie, Narison and Riazuddin~\cite{crai} to deal with the 
electromagnetic mass shift in QCD, that the short distance QCD 
contribution to the ``electroweak" pion mass difference 
away from the chiral limit is ``finite",
contrarily to what is assumed in the literature and to calculate
its value, which is indeed quite small even compared with the
uncertainty in the long distance contribution. 

It is opportune to consider first why the electromagnetic
mass difference of hadrons (in particular, the proton-neutron
mass difference) is supposed to have a divergence,
and how the authors of Ref.~\cite{crai} solved an apparent
puzzle on this subject. Afterwards, considering the prescription
of Ref.~\cite{crai}, the calculation of the electroweak mass
difference of pions follows straightforwardly. The argument
that the electromagnetic proton-neutron mass difference is
divergent is established using the Cottingham formula
for the mass difference~\cite{col}
\begin{equation}
\delta m_h \propto \int \, \frac{d Q^2}{Q^2}
T_{\mu}^{\; \mu} (Q,P) \; ,
\label{cot}
\end{equation}
where $T_{\mu \nu}(q,p)$ is the virtual forward Compton
amplitude for scattering of a virtual photon of momentum
$q$ on a target of momentum $p$ ($Q^2=-q^2$ and $P^2=-p^2$). From the 
operator-product expansion (OPE) we know that the leading operator 
contributing to $T_{\mu}^{\; \mu}$ is $m_f \bar{\psi}_f \psi_f$, where
$m_f$ and $\psi_f$ are quark masses and fields. Since this
quantity is a renormalization group invariant, $T_{\mu}^{\; \mu}$
behaves as a constant at large $q^2$, and the integral diverges.
The puzzle that we referred to above appears when, based on the use of 
Schwinger-Dyson equations, it was affirmed that 
Eq.(\ref{cot}) is finite under certain conditions~\cite{brod},
{\it i.e.} the Compton amplitude would be given by the running
quark masses, and the mass difference reads
\begin{equation}
\delta m_h \propto \int_{Q_0^2}^{\infty} \, \frac{d Q^2}{Q^2}
\, m_f(Q^2) \; ,
\label{cot2}
\end{equation}
where $Q_0^2$ is a cut-off above which we can use OPE and
perturbation theory, and
\begin{equation}
m_f(Q^2) \equiv m_{0f} \left( \alpha_s(Q^2)\right)^{\gamma} \; ,
\label{mrun}
\end{equation}
where $m_f(Q^2)$ is the running mass, $\alpha_s(Q^2)$ is
the strong running coupling constant, and $\gamma= 12/(33-2 n_f)$, where 
$n_f$ is the number of flavors. Eq.(\ref{cot2})
is finite as long as $n_f \geq 11$! The explanation of
these conflicting results was presented in Ref.~\cite{crai}.

The authors of Ref.~\cite{crai} noted that the results of Eq.(\ref{cot})
and Eq.(\ref{cot2}) are different because they depend on the
order of integration of the strong and electromagnetic corrections.
It is obvious that the result should not depend on which interaction
is considered first, and Craigie, Narison and Riazuddin proposed 
an unique prescription that renders the mass shift calculation
independent of the order of integration, as it must be! The
technique named as renormalization group improved perturbation
theory imply in the use of the running quark mass in the 
calculation~\cite{crai}. It was observed that there is nothing
special about the finite result of Eq.(\ref{cot2}) for
$n_f>11$, except that a subset of diagrams in the renormalization
group improved theory are finite, and when $n_f<11$ the regularization 
scheme of Ref.~\cite{crai} for the mass shift gives the
analytic continuation of the finite calculation!
We will not repeat all the details of Ref.~\cite{crai}, and 
in the following we simply use their prescription. At this
point it is almost trivial to foresee our result: the pion
mass difference would diverge similarly to the proton-neutron
mass difference. However, the QCD contribution to the first one 
is quadratic in the running quark masses instead of the linear
dependence in the proton-neutron mass difference. Therefore, 
the integral of the mass shift has a faster
convergence and is finite for $n_f \geq 5$.

To calculate the short distance QCD contribution to 
the electroweak pion mass
difference we follow closely the procedure of Ref.~\cite{mac}.
We start with the propagator $\psi^{(i)}(q^2)$
of the electroweak covariant divergences
of the hadronic currents $A^{\mu \, (i)}$ (where $i=1+\imath 2 \; (3)$
stands for the charged (neutral) axial-vector pion current):
\begin{equation}
\psi^{(i)}(q^2) = \imath \int d^4 \, x e^{\imath q.x}
\langle 0 | T D_\mu A^{\mu \, (i)}(x) D_\nu A^{\nu \, (i)+}(0)|
\rangle \; ,
\label{div}
\end{equation}
which, by PCAC, may also be written as
\begin{equation}
\psi^{(1+\imath 2)}(q^2) \approx \frac{2 f_{\pi^+}^2 m_{\pi^+}^4}
{-q^2 +m_{\pi^+}^2} \; , \;\;\; 
\psi^{(3)}(q^2) \approx \frac{f_{\pi^0}^2 m_{\pi^0}^4}
{-q^2 +m_{\pi^0}^2} \; .
\label{pca}
\end{equation}
Developing Eq.(\ref{div}), and equalizing it to Eq.(\ref{pca})
at $q=0$, we obtain (see Ref.~\cite{mac})
\begin{equation}
2 f_\pi^2 (m_{\pi^+}^2 - m_{\pi^0}^2) \approx e^2 \int
\frac{d^4 \, q}{(2\pi)^4} \left( D_{\mu\nu}^\gamma (q) - D_{\mu\nu}^Z (q) \right)
\left( 2 \Pi_V^{\mu\nu (3)}(q) - \Pi_A^{\mu\nu (1+\imath 2)}(q) \right) \; ,
\label{fp}
\end{equation}
where $D_{\mu\nu}^\gamma \, (D_{\mu\nu}^Z)$ is the photon (weak 
neutral boson) propagator and
$\Pi_{V(A)}^{\mu\nu (i)}$ are the two-point functions of the
vector and axial currents, which can be decomposed as
\begin{equation}
\Pi^{\mu\nu}(q^2)= - (g^{\mu\nu}-\frac{q^\mu q^\nu}{q^2})\Pi^1 (q^2)+
\frac{q^\mu q^\nu}{q^2} \Pi^0 (q^2) \; .
\label{api}
\end{equation}
Eq.(\ref{fp}) contains only the leading contributions to the electroweak 
pion mass difference. We neglected a term originated from the
difference of quark condensates 
$(m_u-m_d) \langle \bar{u} u - \bar{d} d \rangle $, since
the condensates cancel out at leading order. We also neglected
a large part of the weak interaction contribution, which
disappear in the limit $m_u = m_d$~\cite{mac}. This simplifying limit is
assumed throughout our calculation. Finally, the scalar boson terms
have also been dropped out. All these contributions to the 
short distance pion mass difference will be even smaller than the 
one we are considering. 
Inserting Eq.(\ref{api}) into Eq.(\ref{fp}) and working in the
Landau gauge we arrive at
\begin{equation}
2 f_\pi^2 (m_{\pi^+}^2 - m_{\pi^0}^2) \approx 3 \imath e^2
\int \frac{d^4 \, q}{(2\pi)^4} 
\left( \frac{1}{q^2} - \frac{1}{q^2 - M_Z^2} \right)
\left( 2 \Pi_V^{1(3)} (q^2) - \Pi_A^{1(1+ \imath 2)} (q^2) \right) \; .
\label{qua}
\end{equation}

To evaluate the electroweak mass difference of pions we separate
the integral of Eq.(\ref{qua}) into two pieces. From $0$ to $Q_0^2$,
corresponding to the long distance contribution, we can saturate
the $\Pi 's$ by the low-energy resonances $\rho$ and $A_1$, obtaining
the classical result of Das {\it et al.}~\cite{das}. 
All the details of the calculation
up to now can be found in Ref.~\cite{mac}. The remaining
part is the short distance contribution, where the $\Pi 's$ can be
calculated through perturbative QCD, and is the one in which we
are interested. From the results of 
Ref.~\cite{flor} we can determine the QCD prediction
for the difference of the two-point functions
of Eq.(\ref{qua}), which (at dominant order in
the quark mass) is
\begin{equation}
2 \Pi_V^{1(3)} (q^2) - \Pi_A^{1(1+ \imath 2)} (q^2) 
\approx \frac{3}{8 \pi} (m_u+m_d)^2 \; ,
\label{tpf}
\end{equation}
where $m_u$ and $m_d$ are the u and d quark masses. Here we
differ from Ref.~\cite{mac} and consider the prescription of
Craigie, Narison and Riazuddin ~\cite{crai}, introducing the
running and not the bare masses in the calculation of Eq.(\ref{qua}).
We obtain the following integral for the short distance contribution
to the mass difference
\begin{equation}
2 f_\pi^2 (m_{\pi^+}^2 - m_{\pi^0}^2)_{SD} \approx \frac{9}{8\pi} \imath e^2
\int_{Q_0^2}^\infty \frac{d^4 \, q}{(2\pi)^4} 
\frac{-M_Z^2}{q^2 (q^2-M_Z^2)} [m_u(q^2)+m_d(q^2)]^2 \; .
\label{shd}
\end{equation}
In the above expression, if we define $\Delta_{SD} \equiv 
(m_{\pi^+} - m_{\pi^0})_{SD}$, take $m_{\pi^+} + m_{\pi^0} \sim 2 m_\pi$, 
go to the Euclidean space and, as discussed previously, assume 
$m_u = m_d = m_f(Q^2)$, we obtain
\begin{equation}
\Delta_{SD} \approx \frac{9 \alpha}{32 \pi^2}
\frac{M_Z^2}{f_\pi^2 m_\pi} \int_{Q_0^2}^\infty \, dQ^2 \,
\frac{m_f^2(Q^2)}{(Q^2 + M_Z^2)} \; ,
\label{res}
\end{equation}
where $\alpha$ is the electromagnetic coupling constant.
Note that the integral of Eq.(\ref{res}) is similar to the 
ones discussed in Ref.~\cite{crai}, the integrand 
is proportional to $\ln^{-2\gamma}(Q^2)/Q^2$, and $\Delta_{SD}$
is finite as long as $2\gamma > 1$, or $n_f \geq 5$! The fact
that the two-point functions difference in Eq.(\ref{tpf}) cancel 
out at zeroth and first order in the quark masses is fundamental
for this convergence.

The calculation of the integral in Eq.(\ref{res}) is
straightforward~\cite{crai}, and it is necessary to verify that
it does lead to a negligible contribution. Its evaluation
gives the following result
\begin{equation}
\Delta_{SD} \approx - \frac{9 \alpha}{32 \pi^2}
\frac{m_{0f}^2 Q_0^2}{f_\pi^2 m_\pi}
\left( \ln \frac{Q_0^2}{\Lambda_{QCD}^2} \right)^{-2\gamma} \; .
\label{fin}
\end{equation}
Assuming $m_{0f} \sim 10 \, MeV$, and $Q_0^2 \sim 1 \, GeV^2$ as a
good scale above which we can apply perturbative QCD, we obtain
$\Delta_{SD} \approx - 0.02 \, MeV$ which is a value smaller than
the uncertainty present in the long distance contribution to the
pion mass difference, and perfectly negligible. Unfortunately
the short distance behavior of the mass difference $(\Delta m_\pi)$ 
turned out to be highly dependent on the cut-off $Q_0^2$, and 
we must have $\partial \Delta m_\pi / \partial Q_0^2 = 0$,
reflecting the independence of the physical result on the 
choice of this separation scale. However, we can expect that
with a better knowledge of the transition between the
nonperturbative and perturbative regions this abrupt behavior
will be softened.

In conclusion, we have computed the short distance QCD contribution
to the electroweak pion mass difference out of the chiral
limit (assuming $m_u = m_d$), following a prescription
of Craigie, Narison and Riazuddin. We have shown that this 
mass difference is finite and, as expected, is totally negligible
compared to the long distance contribution. Although the
numerical result just serves to corroborate how
negligible it is, we have not seen the
arguments of Ref.~\cite{crai} applied to the pion mass difference
before. Moreover, it is important to stress the effect of
the remaining standard model interactions when discussing
the high energy part of these mass differences, because
the convergence of the result is affected by these interactions.

\acknowledgments    
I am grateful for discussions with P.S.Rodrigues da Silva.                                             
This research was supported in part by the Conselho Nacional de
Desenvolvimento Cient\'{\i}fico e Tecnol\'ogico (CNPq).

\end{document}